\documentclass[aps,prl,twocolumn,superscriptaddress]{revtex4}

\usepackage{graphicx}
\usepackage{amsmath}
\usepackage{amssymb}
\usepackage{mathrsfs}
\usepackage{subfigure}
\usepackage{color}
\usepackage{bm}

\begin{document}

\author{Michael Foss-Feig}
\affiliation{
JILA, NIST, and Department of Physics, University of Colorado, Boulder, CO 80309-0440, USA}

\author{Andrew J. Daley}
\affiliation{
Department of Physics and Astronomy, University of Pittsburgh, Pittsburgh, PA 15260, USA}

\author{James K. Thompson}
\affiliation{
JILA, NIST, and Department of Physics, University of Colorado, Boulder, CO 80309-0440, USA}

\author{Ana Maria Rey}
\affiliation{
JILA, NIST, and Department of Physics, University of Colorado, Boulder, CO 80309-0440, USA}

\title{Steady-state many-body entanglement of hot reactive fermions}

\begin{abstract}
Entanglement is typically created via systematic intervention in the time evolution of
an initially unentangled state, which can be achieved by coherent
control, carefully tailored non-demolition measurements, or
dissipation in the presence of properly engineered reservoirs.  In
this paper we show that two-component Fermi gases at $\sim\mu$K temperatures naturally evolve, in the presence of
\emph{reactive} two-body collisions, into states with highly entangled
(Dicke-type) spin wavefunctions.  The entanglement is a steady-state property that
emerges---without any intervention---from uncorrelated initial
states, and could be used to improve the accuracy of
spectroscopy in experiments with fermionic alkaline earth atoms or fermionic groundstate molecules.
\end{abstract}
\maketitle
\setlength{\parskip}{0pt}

\vspace*{-4mm}

Many-body entangled states are known to be useful
for quantum computing, quantum teleportation and cryptography
protocols \cite{horodecki}, and precision
metrology \cite{giovannetti}.  With these applications as motivation, the
physics community has invested tremendous effort in preparing,
stabilizing, and measuring entangled systems.  Much of this effort has relied on coherent (Hamiltonian) dynamics to arrive at entangled
states starting from less exotic states with only classical correlations.  However,
these approaches typically suffer from the necessity to either carefully
engineer interactions between particles or to prepare extremely pure and
specific initial states (or both).  A bottom up implementation of
coherent control has yielded entangled states of up to
14 atoms with relatively high fidelity \cite{monz} (in ion experiments), and a top down approach has
yielded weakly entangled states in a Bose Einstein condensate of $\sim10^4$ neutral atoms \cite{lucke}.  A promising
alternative to coherent control is the collective-nondemolition measurement of some observable with a finite
variance in an initially classical state.  Such approaches have been
used to generate entanglement (in the form of spin squeezing) amongst
as many as $10^6$ cold thermal atoms \cite{thompson}.  However, collective and
coherence preserving measurements are generically difficult to make, and the induced
non-classical correlations are typically weak.

In this paper, we show that two-component non-degenerate fermionic gases can be
driven by reactive $s$-wave two-body collisions into steady-state spin
configurations that, for a given value of the saturated particle
number, are pure and highly entangled.  The entanglement comes in the
form of Dicke states \cite{dicke}, in which the spin-wavefunction is fully
symmetric under interchange of the particles (with the burden of
fermionic antisymmetry being taken up entirely by the motional degrees
of freedom).  Such states have been sought in experiments with
ultra-cold bosons for use in Heisenberg-limited phase measurements \cite{holland},
however these approaches typically suffer from the necessity to reach
extremely cold temperatures (for the validity of a two-mode
approximation in a double-well potential) or to employ Feshbach resonances \cite{gross2} (to
enhance spin exchange interactions for two-component Bose Einstein condensates).  The only requirements to achieve such entanglement in the steady-state of
lossy non-degenerate fermions are to have an SU(2) invariant single-particle Hamiltonian (in the pseudo-spin degrees of
freedom) and a significant separation of timescales between $s$-wave
and $p$-wave collisions, with the second requirement typically being satisfied for temperatures in or below the
$\mu\mathrm{K}$ range.

Because the desired property (i.e. the Dicke
type of spin-entanglement) persists in the steady-state of dissipative dynamics, we do not rely
on the highly controlled coherent manipulation that is typical of spin-squeezing
experiments with bosons.  Driven, dissipative preparation of nontrivial
steady-states has been considered before in the context of many-body
atomic systems \cite{kraus,diehl, diehl2}, and has been achieved recently in
\cite{barreiro,krauter}.  In contrast to these examples, the mechanism described here is
intrinsic and generic to a variety of interesting and
experimentally relevant systems, such as fermionic alkaline-earth
atoms (AEAs) and fermionic dipolar molecules, and does not require any special
engineering of the system-reservoir coupling.  After presenting calculations in support of
our claims, we discuss the possible realization of such steady
states in an experiment.  In particular, we will propose a simple proof of
principle experiment in which the steady-state entanglement can be
revealed via Ramsey spectroscopy of the $^1S_0$ to $^3P_0$ clock
transition of an AEA \cite{swallows}. In this case, we will see that the interferometric
precision stays relatively constant even as most of the particles are lost
(all but $\sim\sqrt{\mathcal{N}}$ in the long time limit), signaling the
development of quantum correlations and the pursuant violation of the
standard quantum-limit.  The total loss of precision (due to loss of
particles) exactly cancels the gain due to the growth of quantum
correlations.  However, a persistent precision under loss of particles
can provide enhanced spectroscopic accuracy; in particular, the
steady-state under consideration is largely devoid of mean-field clock shifts.

Our description of spin-$\frac{1}{2}$ fermions with two-body reactive
collisions relies on the formalism detailed in Refs. \cite{durr,ripoll,syassen}, generalized for
fermions, where we assume the temperature to be sufficiently low that
losses are dominantly in the $s$-wave channel.  As in
Ref. \cite{durr}, large kinetic energy of fermions in the
outgoing channels (which for reactive molecules can correspond to
temperatures in the $10$K range) guarantees they will be rapidly lost from
any typical atom trap, justifying a Born Markov
approximation.  Given a density matrix $\varrho$ for the system (fermions, Hilbert space $\mathscr{S}$) plus reservoir (outgoing channels of
the inelastic collisions, Hilbert space $\mathscr{R}$), the Born-Markov
approximation leads to a
master equation for the system reduced density matrix
$\rho=\mathrm{Tr}_{\mathscr{R}}[\varrho]$ \cite{ripoll}:
\begin{equation}
\hbar\dot{\rho}=i[\rho,\mathcal{H}]-\frac{\kappa}{2}\int
d^3\bm{r}\left(\mathcal{J}^{\dagger}\mathcal{J}^{}\rho+\rho\mathcal{J}^{\dagger}\mathcal{J}^{}-2\mathcal{J}\rho\mathcal{J}^{\dagger}\right).
\label{meq}
\end{equation}
The system Hamiltonian $\mathcal{H}=\mathcal{H}_0+g\int d^{3}\bm{r}\mathcal{J}^{\dagger}\mathcal{J}^{}$
is composed of an unspecified single-particle term $\mathcal{H}_0$ and
an interaction term with coupling constant $g=4\pi\hbar^2a_R/m$,
where $m$ is the particle mass and $a=a_R+ia_I$ ($a_I<0$) is the
complex $s$-wave scattering length.  The jump operators are defined by
$\mathcal{J}(\bm{r})=\psi^{}_{\uparrow}(\bm{r})\psi^{}_{\downarrow}(\bm{r})$
(their explicit $\bm{r}$ dependence is suppressed in the integrals above),
where $\psi_{\sigma}(\bm{r})$ annihilates a fermion located at
position $\bm{r}$ in internal state $\sigma\in\{\uparrow,\downarrow\}$, and
$\kappa=-4\pi\hbar^2a_I/m$. Assuming without loss of generality
that the initial number of particles $\mathcal{N}$ is even, the
relevant system Hilbert space can be written
as a direct sum over spaces with well defined particle number, $\mathscr{S}=\mathscr{S}^{\mathcal{N}}\oplus\mathscr{S}^{\mathcal{N}-2}\oplus\dots\oplus\mathscr{S}^{0}$, between which coherence never develops.  Hence, the
density matrix can be decomposed into a sum of density matrices in
each particle-number sector, any one of which we label by $\rho^n$
once normalized.  Furthermore, any Hilbert space $\mathscr{S}^n$ can be decomposed into a direct product between motional (m) and spin (s) degrees of
freedom, $\mathscr{S}^n=\mathscr{S}^n_m\otimes\mathscr{S}^{n}_s$, and we
can define a reduced spin density matrix by
$\rho^n_s=\mathrm{Tr}_{\mathscr{S}_m}[\rho^n]$.  For what follows, it will be useful to
define a fidelity in a given Dicke state \footnote{The Dicke state of
  $n$ spins having total $z$ projection of spin $S^z$ is obtained by
  acting on the state with maximal $z$-projection of total spin $(n/2-S^z)$ times with the total
  spin lowering operator $S^{-}$.} of the spin degrees of
freedom of $n$ particles, $|S=n/2,S^z\rangle$, given by the population
of $\rho_s^n$ in the Dicke state
\begin{equation}
\mathcal{F}_{S,S^z}\equiv\langle S,S^z|\rho^n_s|S,S^z\rangle.
\end{equation}
Here $S$ and $S^z$ are quantum numbers for the total spin and its
projection along the $z$-axis, respectively.

\emph{Two particles}. To make the physics clear in a simple context, we begin by considering two fermions in a single
double well potential (which could be formed in an optical
super-lattice \cite{anderlini,trotzky}).  We consider a single wavefunction
$\varphi_{\alpha}(\bm{r})$ in each well ($\alpha\in\{L,R\}$), denote the creation operator for a
fermion in spin state $\sigma$ and wavefunction $\varphi_{\alpha}$ by $\psi^{\dagger}_{\sigma
  \alpha}$, and choose an initial state
$\psi^{\dagger}_{\uparrow L}\psi^{\dagger}_{\downarrow R}|vac\rangle$
without spin correlations.  Within a tight binding model for these two
wavefunctions, the Hamiltonian is
\begin{equation}
\mathcal{H}=-J\sum_{\sigma}(\psi^{\dagger}_{\sigma L}\psi^{}_{\sigma
  R}+\psi^{\dagger}_{\sigma R}\psi^{}_{\sigma
  L})+U\sum_{\alpha=L,R}\mathcal{J}^{\dagger}_{\alpha}\mathcal{J}^{}_{\alpha},\end{equation}
and the master equation reads
\begin{equation}
\hbar\dot{\rho}=i[\rho,\mathcal{H}]-\frac{\gamma}{2}\sum_{\alpha=L,R}\left(\mathcal{J}^{\dagger}_{\alpha}\mathcal{J}^{}_{\alpha}\rho+\rho \mathcal{J}^{\dagger}_{\alpha}\mathcal{J}^{}_{\alpha}-2\mathcal{J}^{}_{\alpha}\rho\mathcal{J}^{\dagger}_{\alpha}\right).
\label{meq2well}
\end{equation}
\begin{figure}[!t]
\centering
\includegraphics[width=7.0 cm]{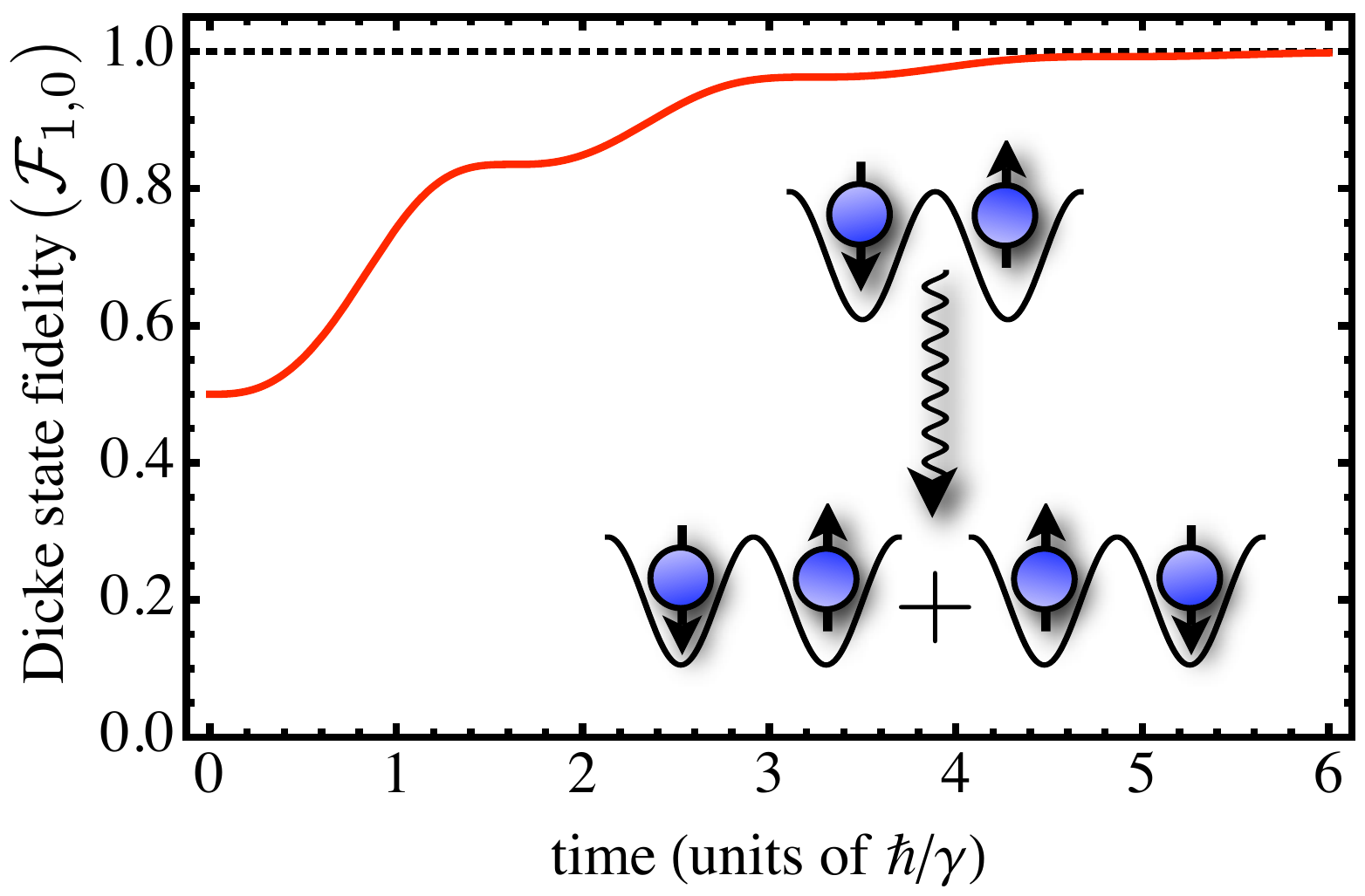}
\caption{(Color online).  The solid red line is the
  fidelity of the final density matrix (after post-selection for a non-vacant well) with respect to the $S_z=0$ Dicke state, and approaches 1
  (black dotted line) at times long compared to $\gamma^{-1}$.  The oscillations imposed over the exponential decay are due to the
  inter-well hopping.}
\label{fig1}
\end{figure}
Here $J$ is the inter-well hopping, $\mathcal{J}_{\alpha}=\psi^{}_{\uparrow \alpha}\psi^{}_{\downarrow
  \alpha}$, $U=g\int
d^3\bm{r} |\varphi_{\alpha}(\bm{r})|^4$ is the onsite interaction energy, and
$\gamma=\kappa\int d^{3}\bm{r}|\varphi_{\alpha}(\bm{r})|^4$ is the
onsite loss rate.  The initial state can be decomposed into an evenly
weighted superposition of triplet and singlet [$(\psi^{\dagger}_{\uparrow
  L}\psi^{\dagger}_{\downarrow R}\pm\psi^{\dagger}_{\downarrow
  L}\psi^{\dagger}_{\uparrow R})|vac\rangle$, with plus for the
triplet], and the spin wave function of the triplet is the entangled Dicke state
$|1,0\rangle$.  The triplet, $|t\rangle$, having a spin wavefunction that is symmetric
under exchange, has an orbital wavefunction that is antisymmetric
under exchange, and therefore it is ``dark'' to $s$-wave losses, by which
we mean simply that $\mathcal{J}_{L}|t\rangle=\mathcal{J}_{R}|t\rangle=0$.  It also
happens to be an eigenstate of $\mathcal{H}$, and so it is stationary
under propagation of the master equation (\ref{meq2well}).  On the other
hand, there are no dark eigenstates in the singlet sector, and as a
result $\rho^2_s$ is pure at long times and satisfies
$\mathcal{F}_{1,0}=1$.  In other words the steady state of our
system, when restricted to the subspace with two particles,
is the entangled Dicke state $|1,0\rangle$.  It should be noted that
there is also a $50\%$ probability of obtaining the vacuum, and hence
in an array of double wells the entanglement fidelity is only unity
after post-selection of the non-vacant wells.  In this simple
example we see an important general feature of the physics we will
discuss, that even purely \emph{local} (intra-well) dissipation, when
coexisting with Hamiltonian dynamics that delocalizes the particles, generates
\emph{non-local} (inter-well) spin correlations in the steady state.

\emph{Many particles}.  Solving Eq. (\ref{meq}) exactly for
initial states with $\mathcal{N}>2$ initial particles quickly becomes
impossible, but strong statements can nevertheless be made regarding
the steady state.  It is crucial to observe that the jump operators only
remove spin singlets from the system, which follows from Fermi
statistics combined with the even exchange symmetry of the spatial part of any two-particle wave
function susceptible to $s$-wave scattering.  Intuitively, this
suggests that losses do not decrease the expectation value of the
total spin, $\bm{S}=\frac{1}{2}\int
d^3\bm{r}\psi^{\dagger}_{\sigma}(\bm{r})\bm{\tau}_{\sigma\sigma'}\psi^{}_{\sigma'}(\bm{r})$
($\bm{\tau}$ being a vector whose components are the Pauli matrices).  Mathematically, we say that $\frac{d}{dt}\langle\bm{S}\cdot\bm{S}\rangle=\mathrm{Tr}[\rho\bm{S}\cdot\bm{S}]=0$,
which can easily be verified in the case when $\mathcal{H}$ is SU$(2)$
invariant by checking that
$[\bm{S}\cdot\bm{S},\mathcal{J}(\bm{r})]=0$.  A stronger consequence of
the commutation of all $\mathcal{J}(\bm{r})$ with $\bm{S}\cdot\bm{S}$ is that
population in any sector of total spin,
$\mathcal{P}_S$, is also conserved.  Because any state with well
defined total spin $S$ must have $\langle \hat{N}\rangle\geq2S$
particles (where $\hat{N}=\int
d^3\bm{r}\psi^{\dagger}_{\sigma}\psi^{}_{\sigma}$ is the total number operator), an
immediate consequence is that the loss of particles can only yield the
vacuum deterministically at long times if the initial state is a total spin-singlet.  For
an uncorrelated spin state, such as a non-degenerate thermal
distribution of $\mathcal{N}$ fermions in a balanced incoherent mixture of
$\uparrow$ and $\downarrow$, it can be shown that \cite{arecchi}
\begin{equation}
N(t)\equiv\mathrm{Tr}[\rho\hat{N}]\geq\sum_S
2S\mathcal{P}_S=\frac{\pi^{1/2}\Gamma\left[\frac{\mathcal{N}}{2}+1\right]}{\Gamma[\frac{\mathcal{N}}{2}+\frac{1}{2}]}-1,
\label{bound}
\end{equation}
which places a lower bound on the steady-state expectation value for
the number of particles $N(t)$.  This expectation value determines the particle
number in a typical steady-state configuration, and is achieved (on
average) \emph{without} any post selection, but variations of the steady-state
particle number will occur from shot to shot.  Taking Stirling's approximation for large
$\mathcal{N}$ yields an approximate bound
$N(t)\gtrsim\sqrt{\pi\mathcal{N}/2}$.  For the chosen restriction on the initial state, the validity of
Eq. (\ref{bound}) depends only on the SU($2$) invariance of $\mathcal{H}$, and not on its precise form.
Whether the bound (\ref{bound}) is saturated in the steady-state,
however, is an important and delicate issue; an affirmative answer
guaranties that all of the $\rho^n_s$ describe pure Dicke states in the
steady-state.  Demonstrating that this bound is indeed saturated in
certain experimentally relevant situations, namely a 1D harmonic trap
and a 1D Hubbard chain (optical lattice), is a central technical result of this paper.

\begin{figure}[!t]
\centering
\includegraphics[width=8 cm]{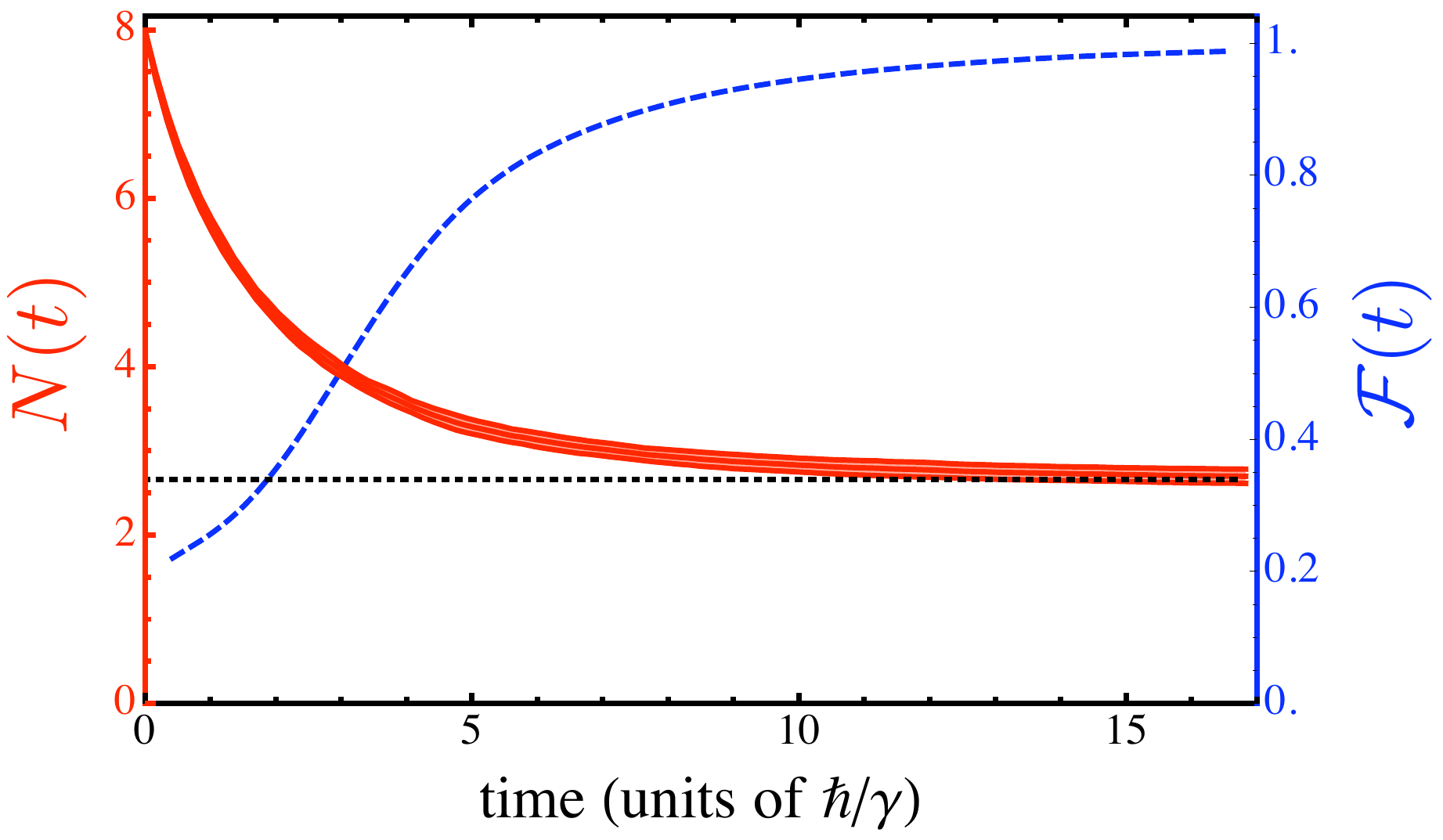}
\caption{(Color online). Calculation of particle number [$N(t)$, solid red
  line] and average Dicke state fidelity [$\mathcal{F}(t)$, dashed blue line] for an 8-site Hubbard chain via quantum
  trajectories. For the number, the shaded region is an estimate of the
  statistical error from sampling of a finite number of trajectories.
The black dotted line is the analytic bound in Eq. (\ref{bound}).}
\label{fig2}
\end{figure}

Saturation of the bound in Eq. (\ref{bound}) is guaranteed if, for
any fixed value of $n$ and $S_z$, the pure density matrix
$|n/2,S_z\rangle\langle n/2,S_z|$ is
the \emph{unique} steady-state reduced spin density matrix.  This uniqueness, in
turn, is equivalent to requiring that any dark-state with quantum
numbers $n$ and $S_z$ has a well defined spin wavefunction given by the Dicke state $|n/2,S_z\rangle$.  In the supplement we prove
this to be true for a 1D harmonic oscillator potential, and we have
verified it numerically for a 1D Hubbard chain (see below).  It is worth noting at this point that, while the equivalence of dark-states
with the Dicke states is intuitive, there are natural Hamiltonians for
which this intuition is incorrect.  In particular, all Hamiltonians in $D>1$ that are
separable in cartesian coordinates do have dark-states with $\mathcal{F}_{S,S_z}<1$.

In order to verify the above statements numerically, we have
performed quantum trajectories simulations for an $8$-site Hubbard chain with open boundary
conditions, an initial filling of
one particle per site, and zero polarization ($\mathcal{N}=8$ and
$S_z=0$).  In general we have lower-densities in mind for any
experimental application, but using one particle per site allows us to
stretch the numerics to the largest $\mathcal{N}$ possible.  In Fig. \ref{fig2} we show the calculated
particle number and average Dicke state fidelity, $\mathcal{F}(t)=\frac{1}{4}\sum_{S=1}^{4}\mathcal{F}_{S,0}$, and one can see that the
former saturates the bound Eq. (\ref{bound}) while the latter approach
unity at long times.  For this calculation we solve for $\mathcal{O}(10^4)$
trajectories with no approximation.

\emph{Experimental realization}.  Dicke states are known to be useful for a variety of quantum
information protocols, including but not limited to quantum secret
sharing \cite{prevedel}, teleportation \cite{kiesel}, and sub shot-noise limited precision
spectroscopy \cite{holland}.  Here we give a brief description of how the
proposed Dicke state preparation could be used in precision
spectroscopy of the clock transition in alkaline-earth atoms.
For a fixed interrogation time, spectroscopy on $\mathcal{N}$ uncorrelated atoms has a phase
sensitivity $\delta\varphi\gtrsim1/\sqrt{\mathcal{N}}$, a bound known
as the standard quantum limit (SQL).  This bound can be understood as
the minimum tipping angle needed to cause a coherent spin state (CSS)
to have an uncertainty cone that precludes its initial position
\cite{kasevich} (to one standard deviation).  On the other hand, spectroscopy
on a Dicke state of $\mathcal{N}$ particles with spin $S_z=0$ has the potential to
reach the Heisenberg limit (HL) of phase sensitivity,
$\delta\varphi\sim1/\mathcal{N}$ \cite{holland,kasevich}.  It is important to realize that the production of Dicke states with
$\sqrt{\mathcal{N}}$ fermions via two-particle loss does not actually enhance the phase sensitivity
relative to the initial state with $\mathcal{N}$ fermions; the
enhancement in phase sensitivity between the SQL and HL exactly
compensates the reduction of particle number.  However, the reduced
particle number in the Dicke state and darkness to \emph{real}
$s$-wave interactions (which if present generate clock shifts), can
render the accuracy of the final Dicke state superior to that of the initial $\mathcal{N}$ fermion uncorrelated state.

Rather than allowing losses amongst a macroscopic sample of atoms, for
which the approach to the steady state could be quite slow, we imagine an
array of $\mathcal{T}$ 1D tubes created by a 2D optical lattice.
Although there will be variations in the atom number from tube to
tube, for simplicity we take each tube to have exactly $\mathcal{N}$
fermionic AEAs in the $^{1}S_0$ electronic state and
$I^z=I$ nuclear-spin state, denoted $|^1S_0,I\rangle$.  For the analysis in this paper to be
valid, the temperature should be small compared to the vibrational
level-spacing in the transverse tube direction, and also low enough that only the harmonic
part of the trapping potential along the tube axis is sampled by the
atoms.  A $\pi/2$-pulse on the spin degrees of freedom [$|^1S_0,I\rangle\rightarrow\frac{1}{\sqrt{2}}(|^1S_0,I\rangle+|^1S_0,I-1\rangle)$], followed by single particle dephasing \footnote{Dephasing of
  nuclear spins is typically extremely slow, but can be briefly
  enhanced with large magnetic field gradients or by scattering photons on a cycling
transition.}, generates
a statistical mixture of the two spin states ($I^z=I$ and $I^z=I-1$).  Losses can be initiated
by applying a $\pi$-pulse on the clock transition ($|^1S_0,I^z\rangle\rightarrow|^3P_0,I^z\rangle$).  We estimate that this $\pi$-pulse \footnote{This transfer
  needs to be of high fidelity, since any atoms remaining in the
  $^1S_0$ state can undergo $s$-wave collisions with $^3P_0$ atoms that
  \emph{do} reduce the system spin.  In principle, the purity of this
  transfer could be enhanced by using an optical pumping scheme.} can be achieved on the $\lesssim100\mu$s timescale
without exciting transverse excitations in the tubes (which, if
present, violate the assumption of a 1D geometry and destroy the
uniqueness of the steady-state).  Thus the transfer into $^3P_0$ is sufficiently fast that it can be considered instantaneous on the initial
timescale of reactive collisions---which, based on universal
considerations for a Lieb-Liniger gas, we estimate to be $\gtrsim1$ms
for experimentally relevant 1D densities \cite{durr}---such that it suddenly initiates strong $2$-body
$s$-wave losses.

The steady state of the system is a statistical mixture of Dicke states
in the different tubes, each having some value of $\mathcal{D}_j$
particles (centered around $\mathcal{D}_0\approx\sqrt{\mathcal{N}}$)
and spin projection $S^z_j$ (centered around zero).  Spin selective
transfer of $|^3P_0,I-1\rangle$ into $|^1S_0,I\rangle$ maps the spin degree of freedom onto
the clock states, leaving a spin-polarized sample, and Ramsey spectroscopy on the
clock transition can then be performed \cite{kasevich}.  Despite the
fluctuation of both $\mathcal{D}_j$ and $S^z_j$ from one tube to another,
it can be shown (see the supplement) that the minimum resolvable
rotation angle in a Ramsey experiment scales as
\begin{equation}
\delta\varphi_{\mathrm{min}}\sim1/\mathcal{D}_0\sqrt{\mathcal{T}}.
\label{sensitivity}
\end{equation}
This result can be interpreted as the existence of Heisenberg limited sensitivity for each tube, which is
then combined between tubes in a statistically independent manner
(hence the $1/\sqrt{\mathcal{T}}$).  In order to utilize this phase
sensitivity, the initial value of $S^z=\sum_{j}S^z_j$ for the entire ensemble must be
accurately known.  Because $S^z$ is conserved by the losses, it can be
measured before transfer to the $^3P_0$, and hence the measurement does not need to
preserve any inter-particle correlations (since these develop during
the losses).  Accurate measurements of this type and precision for
$\sim100$ atoms in an optical cavity have recently been demonstrated \cite{vuletic}.

\begin{figure}[!t]
\centering
\includegraphics[width=8.5 cm]{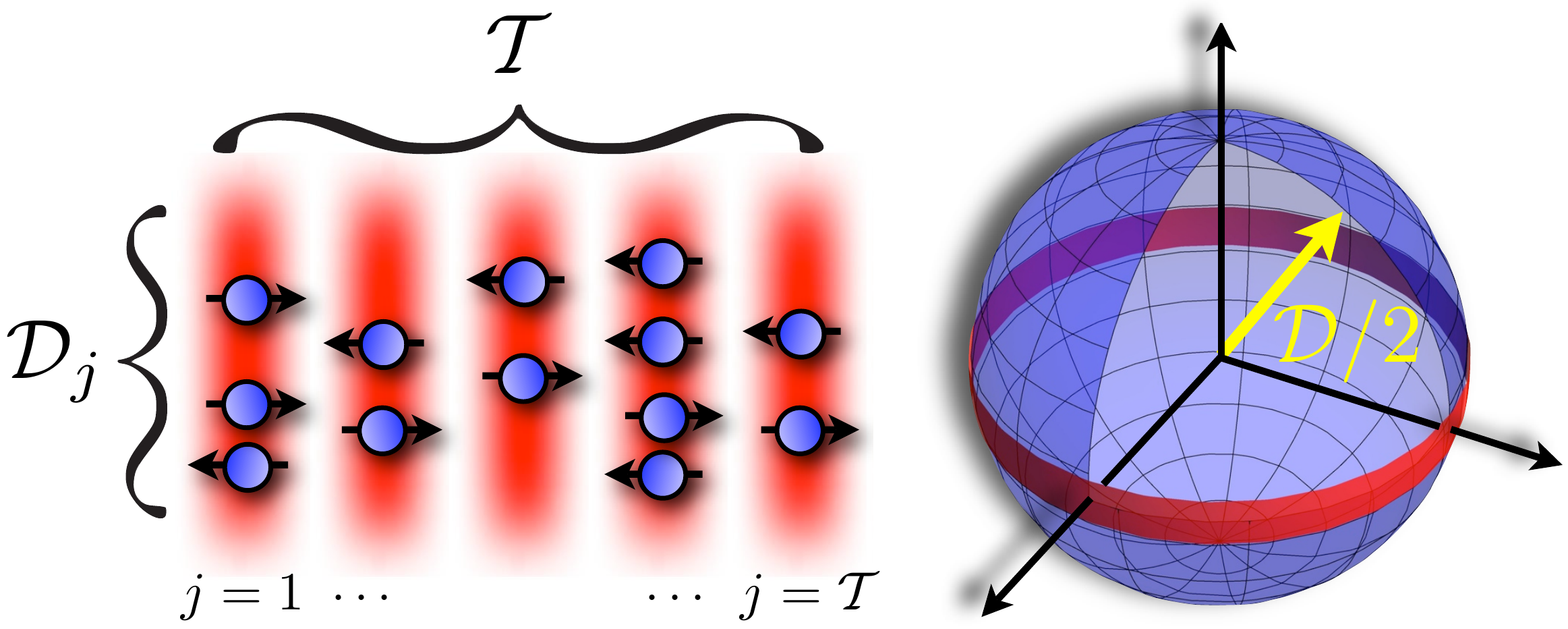}
\caption{(Color online). (a) An array of $\mathcal{T}$ 1D tubes, each
  having $\mathcal{D}_j$ atoms in a Dicke state. (b)
  Bloch sphere representation of a Dicke state in a particular tube. }
\label{fig3}
\end{figure}

The primary limitations on the final state fidelity achievable in
experiments is likely to be a combination of finite $p$-wave losses
(which the Dicke states are not dark to) and magnetic field gradients.
At sub $\mu\mathrm{K}$ temperatures, the $s$-wave losses in a spin mixture of
$^{87}$Sr are expected to be about an order of magnitude faster than
the $p$-wave losses \cite{bishof}.  For reactive molecules (or
$^{171}\mathrm{Yb}$), where the inelastic collisions are expected to more fully saturate the
unitarity bound \cite{goulven}, this separation of rates will most
likely be even larger.  Magnetic field gradients couple sectors
of different total $S$, all of which are separated from the Dicke
manifold by a gap for finite systems and nonzero $a_{R}$, so in principle their adverse effects can
be suppressed to first order \cite{rey}.  Furthermore, if the
two components of the Fermi gas are two nuclear spin states of an
AEA, they will be extremely insensitive to magnetic
field gradients: We estimate that typical gradients ($1\mathrm{mG}/\mathrm{cm}$) will cause spin
dephasing on a $100\mathrm{s}$ timescale for a linear system size of
$100\mu\mathrm{m}$.  This timescale is several orders of magnitude longer than
the initial two-body loss rate in tightly confined 1D tubes, which we estimate to
be on the order of $10$ms for $^{87}$Sr (assuming a
$50E_\mathrm{R}$ 2D lattice and scaling the density dependent
loss rate from Ref. \cite{bishof}), and even faster for $^{171}$Yb \cite{goulven}.  A more quantitative analysis of the effects of
both magnetic field imperfections and finite $p$-wave losses requires
numerical simulations beyond the scope of this work, and is left for
future study.

\emph{Conclusions}.  In this paper we have demonstrated that fairly unrestrictive
initial conditions, without intervention or engineering and in the
presence of reactive two-body collisions, are sufficient to generate
steady-state spin entanglement between non-degenerate fermions.  These reactive collisions, which occur both
in optically excited alkaline earth atoms and many dipolar molecules
(e.g. KRb), are typically viewed as an impediment to interesting
physics, but clearly this need not be the case.  We expect this
physics to enable the distillation of Dicke states from initially
uncorrelated fermionic atoms and molecules, hence extending the scope
of a variety of experimental progress made in the spin squeezing of bosons.

We thank Jun Ye, Kaden Hazzard, and Goulven Qu\'em\'ener for
helpful discussions.  This work was supported by NIST, the NSF (PIF and
PFC grants), AFOSR and ARO individual investigator awards, and the ARO with funding from the DARPA-OLE program.

\setcounter{equation}{0}
\renewcommand{\theequation}{S\arabic{equation}}
\appendix
\section{Uniqueness of the steady state \label{a1}}
Here we show that for a 1D harmonic oscillator, in a particular sector of Hilbert space
$\mathscr{S}^n$ and for a particular value of $S^z$, the unique steady-state reduced spin density matrix
is given by
\begin{equation}
\rho_s^n=|n/2,S^z\rangle\langle n/2,S^z|.
\end{equation}
The extension of what follows to the 1D Hubbard chain is fairly
straightforward, and will be described in more detail in future work.
As discussed in the text, it is sufficient to prove that all dark eigenstates of the non-interacting Hamiltonian
\begin{equation}
\mathcal{H}_{HO}=\int
dx\;\psi^{\dagger}_{\sigma}(x)\left(\frac{\partial_x^2}{2m}+\frac{1}{2}m\omega^2
x^2\right)\psi^{}_{\sigma}(x)
\end{equation}
have a maximally symmetric spin wavefunction.  To understand the properties of its dark eigenstates under particle exchange, we will actually work in first quantization writing an eigenstate for $\mathcal{N}$ particles as
\begin{equation}
\Psi=\sum_{\vec{\sigma}}\mathcal{A}_{\vec{\sigma}}\;\Phi_{\vec{\sigma}}(r_1,\dots,r_{\mathcal{N}})|\vec{\sigma}\rangle.
\end{equation}
Here the $j^{\mathrm{th}}$ component of the vector $\vec{\sigma}$, $\sigma_j\in\{\uparrow,\downarrow\}$, labels the spin orientation of the $j^{\mathrm{th}}$ particle (along some arbitrary quantization axis, which we'll call $z$), and the total spin wavefunction in any term of the sum is
\begin{equation}
|\vec{\sigma}\rangle=|\sigma_1\rangle\otimes|\sigma_2\rangle\otimes\dots\otimes|\sigma_{\mathcal{N}}\rangle.
\end{equation}
The sum over $\vec{\sigma}$ should be understood as independent summations over each index
\begin{equation}
\sum_{\vec{\sigma}}=\sum_{\sigma_1}\sum_{\sigma_2}\cdots\sum_{\sigma_\mathcal{N}},
\end{equation}
the coefficients $\mathcal{A}_{\vec{\sigma}}$ are arbitrary, and
$\Phi_{\vec{\sigma}}$ is a normalized orbital wavefunction for the
$\mathcal{N}$ particles.  Dark states of $s$-wave losses have zero
expectation value in the interaction operator
\begin{equation}
\mathcal{U}=\sum_{m<n}\mathcal{U}_{mn}=g\sum_{m<n}\delta(r_m-r_n),
\end{equation}
and this expectation value can be evaluated as
\begin{eqnarray}
U&=&\int\mathscr{D}r\;\Psi^{*}\,\mathcal{U}\,\Psi \nonumber \\
&=&\sum_{\vec{\sigma}}|\mathcal{A}_{\vec{\sigma}}|^2\int\mathscr{D}r\;\Phi^{*}_{\vec{\sigma}}\,\mathcal{U}\,\Phi_{\vec{\sigma}},
\end{eqnarray}
with $\mathscr{D}r\equiv\prod_{j}dr_j$.  The last equality holds because the interaction is spin-independent.  It is crucial to realize that the operator $\mathcal{U}$ is positive-semidefinite, which means that satisfying $U=0$ actually implies the stricter constraint
\begin{equation}
\int\mathscr{D}r\;\Phi^{*}_{\vec{\sigma}}\,\mathcal{U}\,\Phi_{\vec{\sigma}}=0~~~~\forall\vec{\sigma}.
\end{equation}
 In addition to the operator $\mathcal{U}$ being positive-semidefinite, the constituent pairwise interaction operators are as well.  Hence, the condition $\int\mathscr{D}r\;\Phi^{*}_{\vec{\sigma}}\,\mathcal{U}\,\Phi_{\vec{\sigma}}=0$ actually implies that
\begin{equation}
\int\mathscr{D}r\;\Phi^{*}_{\vec{\sigma}}\,\mathcal{U}_{mn}\,\Phi_{\vec{\sigma}}=0~~~\forall\;m\neq n.
\end{equation}
The above set of equalities can now be used to pin down properties regarding the exchange symmetry of the wave functions $\Phi_{\vec{\sigma}}$.

\subsubsection{Implications for the exchange symmetry of the orbital wave functions}

\noindent Let's choose two particles, say particle 1 and particle 2, and define
relative and center-of-mass coordinates for them as $r=r_1-r_2$ and
$R=(r_{1}+r_{2})/2$.  We can then expand the wavefunction
$\Phi_{\vec{\sigma}}$ in a way that makes the symmetrization with
respect to exchange of particles 1 and 2 explicit:
\begin{equation}
\Phi_{\vec{\sigma}}=\sum_{\beta s}\mathcal{B}_{\vec{\sigma}\beta s}\times\varphi_{\alpha(\beta,s)}(r)\varphi_{\beta}(R)\psi_s(r_3,\dots,r_N).
\end{equation}
In the above $\varphi_{\alpha}$ are harmonic oscillator wavefunctions, the $\psi_s$ are a complete and orthonormal set of eigenfunctions for the remaining $\mathcal{N}-2$ particles, and the notation $\alpha(\beta,s)$ implies that the relative wavefunction of particles 1 and 2 is uniquely determined by $\beta$ and $s$.  This point is crucial, and relies on the observation that for $\Psi$ to be an eigenstate, when expanded in terms of eigenstates $\Phi_{\sigma}$ all of the eigenstates must have the same eigenvalue.  Hence the energies of states $\alpha$, $\beta$, and $s$ are constrained to add to some fixed value.  From now on we'll drop this explicit dependence.  The interaction energy between particles 1 and 2 is given by
\begin{eqnarray}
U_{12}&=&\int\mathscr{D}r\;\Phi^{*}_{\vec{\sigma}}\,\mathcal{U}_{12}\,\Phi_{\vec{\sigma}}
\nonumber \\
&=&g\sum_{\beta s}\mathcal{B}_{\vec{\sigma}\beta
  s}\varphi_{\alpha}(0)\times\mathcal{B}^{*}_{\vec{\sigma}\beta
  s}\varphi^{*}_{\alpha}(0)
\nonumber  \\
&=&\sum_{\beta s}\left|\mathcal{B}_{\vec{\sigma}\beta
  s}\varphi_{\alpha}(0)\right|^2,
\end{eqnarray}
and hence $U_{12}=0$ implies that
\begin{equation}
\mathcal{B}_{\vec{\sigma}\beta
  s}\varphi_{\alpha}(0)=0~~~\forall~\beta,s
\end{equation}
So $\mathcal{B}_{\vec{\sigma}\beta s}$ must be zero for all
even wave functions (all of which are finite at the origin), implying
that $\Phi_{\vec{\sigma}}$ is strictly odd under
interchange of particles 1 and 2.  By repeating the above argument for
two arbitrary particles $m$ and $n$, it is easy to see that
$\Phi_{\vec{\sigma}}$ is strictly odd under interchange of \emph{any}
two particles.

\subsubsection{Implications for the spin wavefunction}

We can now ask what the antisymmetry of $\Phi_{\vec{\sigma}}$ implies
for the full wave function
\begin{equation}
\Psi=\sum_{\vec{\sigma}}\mathcal{A}_{\vec{\sigma}}\;\Phi_{\vec{\sigma}}(r_1,\dots,r_N)|\vec{\sigma}\rangle.
\end{equation}
Under interchange of two arbitrary particles we have
$|\vec{\sigma}\rangle\rightarrow|\vec{\sigma}'\rangle$, and we obtain the new wave
function
\begin{eqnarray}
\Psi'&=&-\sum_{\vec{\sigma}}\mathcal{A}_{\vec{\sigma}}\;\Phi_{\vec{\sigma}}(r_1,\dots,r_N)|\vec{\sigma}'\rangle
\nonumber \\
&=&-\sum_{\vec{\sigma}}\mathcal{A}_{\vec{\sigma}'}\;\Phi_{\vec{\sigma}'}(r_1,\dots,r_N)|\vec{\sigma}\rangle
\nonumber \\
&=&-\Psi \nonumber \\
&=&-\sum_{\vec{\sigma}}\mathcal{A}_{\vec{\sigma}}\;\Phi_{\vec{\sigma}}(r_1,\dots,r_N)|\vec{\sigma}\rangle,
\end{eqnarray}
implying that
\begin{equation}
\mathcal{A}_{\vec{\sigma}}\;\Phi_{\vec{\sigma}}(r_1,\dots,r_N)=\mathcal{A}_{\vec{\sigma}'}\;\Phi_{\vec{\sigma}'}(r_1,\dots,r_N).
\end{equation}
The second equality follows because switching
$\sigma\leftrightarrow\sigma'$ in the summand just changes the order
of the terms in the sum, and the third equality follows from
the antisymmetry of the total wavefunction under particle exchange.  By repeated permutations of various particles, this chain of logic can
be used to demonstrate that all of the
$\mathcal{A}_{\vec{\sigma}}\;\Phi_{\vec{\sigma}}(r_1,\dots,r_N)$ are
equal, and hence we have
\begin{equation}
\Psi=\mathcal{A}\Phi(r_1,\dots,r_N)\sum_{\vec{\sigma}}|\vec{\sigma}\rangle.
\end{equation}
Now $\mathcal{A}$ is just some normalization, which is related to the
total $z$ projection of the spin, and it is clear that $\Psi$ breaks
up into the product of a completely antisymmetric orbital wavefunction
multiplied by a completely symmetric spin wavefunction.

\section{Phase sensitivity of the steady-state \label{a2}}

In order to estimate the phase measurement sensitivity of an array of 1D tubes in
the steady state, we begin by considering just the $j^{\mathrm{th}}$ tube, with
initial particle number $\mathcal{N}$, final particle number
$\mathcal{D}_j$ after relaxing to steady state via collisional loss,
and initial (and final) spin projection $S^z_j$.  The initial
$\mathcal{N}$ atoms can most easily be prepared in an incoherent
mixture of spin up and spin down by simply allowing a coherent state
initially prepared along the $x$-direction to undergo single particle
dephasing (which could be briefly enhanced via a myriad of methods).  For the coherent state, the probability of a given $S^z_j$ is given by a binomial distribution, which for large $\mathcal{N}$ is approximated by the continuous probability distribution
\begin{equation}
\mathcal{P}(S^z_j)=\sqrt{\frac{2}{\mathcal{N}\pi}}e^{-2(S^z_j/\sqrt{\mathcal{N}})^2}.
\end{equation}
Note that without dephasing into a mixture, a coherent state of
fermions does not undergo $s$-wave collisions.  In the sense that such a distribution is easily prepared
experimentally, we take this to be a worst-case scenario; a
distribution of $S^z_j$ more sharply peaked around $S^z_j=0$ will enhance the phase sensitivity.  The
probability distribution of steady-state particle numbers in the $j^{\mathrm{th}}$ tube,
conditioned on a particular value of $S^z_j$, is given by
\begin{equation}
\mathcal{P}(\mathcal{D}_j|S^z_j)=\Theta(\mathcal{D}_j-|2S^z_j|)\frac{2\mathcal{D}_j}{\mathcal{N}}e^{-(\mathcal{D}_j/\sqrt{2\mathcal{N}})^2}e^{2(S^z_j/\sqrt{\mathcal{N}})^2}.
\end{equation}
For $S^z_j=0$, this distribution is peaked around $\mathcal{D}_0\approx\sqrt{\mathcal{N}}$ (giving
the expected value of $n(\infty)$ quoted in the text).  For $|S^z_j|>0$, the step function $\Theta$ reflects the fact that
as particles are lost (remember that $S^z_j$ is conserved by
the losses), the remaining particles are maximally spin polarized once
$\mathcal{D}_j=2|S^z_j|$.  The second exponential provides the proper
normalization, $\frac{1}{2}\int_{0}^{\infty}
d\mathcal{D}_j\;\mathcal{P}(\mathcal{D}_j|S^z_j)=1$, where the factor of
$\frac{1}{2}$ comes from converting sums into integrals while respecting our
assumption of even particle number.

For this single tube, small rotations about the $x$-axis by an angle
$\delta\varphi$ cause a standard deviation in the final distribution of
$S^z_j$ given by \cite{kasevich}
\begin{equation}
\sigma(\delta\varphi,\mathcal{D}_j,S^z_j)\approx\delta\varphi\sqrt{\frac{(\mathcal{D}_j-2S^z_j)(\mathcal{D}_j+2S^z_j)}{8}}.
\end{equation}
for large $\mathcal{D}_j$.  For $S^z_j=0$ (before the rotation), this
 demonstrates that a discrepancy in $S^z_j$ (after the rotation) of order
unity is expected for $\delta\varphi\sim1/\mathcal{D}_j$, hence the
Heisenberg limited phase-sensitivity within a single tube.

Our estimation of the phase sensitivity for an array of tubes relies
only on the assumption that the \emph{initial} (i.e. before the losses) value of
$S^z=\sum_jS^z_j$ is known to within an uncertainty $\Sigma$, but does not require any
knowledge of $S^z_j$ in the individual tubes, greatly relaxing
the experimental requirements.  This uncertainty $\Sigma$ guarantees that, in
principle, rotations causing deviations in $S^z$ of order $\Sigma$ can be detected.  
For $\mathcal{T}$ tubes with well defined (i.e. measured) total spin projection $S^z$, the expected standard
deviation in $S^z$ (denoted $\sigma_{\mathrm{tot}}$) due to a rotation by angle $\varphi$ about the
$x$-axis satisfies
\begin{widetext}
\begin{equation}
\sigma_{\mathrm{tot}}^2=\overbrace{\int_{-\infty}^{\infty}\dots\int_{-\infty}^{\infty}}^{\mathcal{T}~\mathrm{times}}\prod_{j}dS^z_j\overbrace{\int_{0}^{\infty}\dots\int_0^{\infty}}^{\mathcal{T}~\mathrm{times}}\prod_{j}d\mathcal{D}_j\prod_{j}\left[\frac{1}{2}\mathcal{P}(\mathcal{D}_j|S^z_j)\mathcal{P}(S^z_j)\right]\,\sum_{j}\sigma(\delta\varphi,\mathcal{D}_j,S^z_j)^2\;\delta(S^z-\sum_{j}S^z_j).
\end{equation}
\end{widetext}
Here $\delta$ is the Dirac $\delta$-function, reflecting the
correlations established between the various $S^z_j$ by the knowledge
of $S^z$, and the factors of $\frac{1}{2}$ again come from converting sums
into integrals while respecting the assumption of even particle
number.  If we ignore this $\delta$-function constraint, which is
valid in the large $\mathcal{T}$ limit, the
integral simplifies greatly to
\begin{equation}
\sigma_{\mathrm{tot}}^2\approx\frac{\mathcal{T}}{2}\int_{-\infty}^{\infty}dS^z_j\int_{0}^{\infty}
d\mathcal{D}_j\mathcal{P}(\mathcal{D}_j|S^z_j)\mathcal{P}(S^z_j)\,\sigma(\delta\varphi,\mathcal{D}_j,S^z_j)^2.
\label{STotVar}
\end{equation}
Equation (\ref{STotVar}) can be evaluated explicitly to reveal
\begin{equation}
4\sigma_{\mathrm{tot}}^2\approx\mathcal{T}\mathcal{N}\delta\varphi^2\approx\mathcal{T}\mathcal{D}_0^2\delta\varphi^2.
\end{equation}
Setting the total standard deviation to $\Sigma$ gives a minimum phase
sensitivity of
\begin{equation}
\delta\varphi_{\mathrm{min}}\approx\frac{2\Sigma}{\mathcal{D}_0\sqrt{\mathcal{T}}}.
\end{equation}
As has been demonstrated recently in Ref. \cite{vuletic},
$\Sigma\sim1$ is possible for $\sim100$ atoms in an optical cavity, as long as
measurements that do not preserve coherence between the atoms are
acceptable.  Such measurements certainly are acceptable before the losses take
place, since we require no inter-particle correlations in the initial
state (they develop dynamically due to the losses).  Therefore, as
quoted in the manuscript, we expect a minimum phase sensitivity of $\delta\varphi_{\mathrm{min}}\sim\frac{1}{\mathcal{D}_0\sqrt{\mathcal{T}}}$
to be achievable in experiment.

\end{document}